\documentclass[aps,prl,showpacs,floatfix,twocolumn,amsmath,amssymb,nofootinbib]{revtex4-1}

\usepackage{amsmath,amsthm,amssymb,color,psfrag,hyperref}
\usepackage{latexsym}
\usepackage{natbib}
\usepackage{graphicx}

\newcommand{\aj}{Astron. J.}
\newcommand{\aap}{Astron. \& Astro.}

\begin{document}

\title{Testing Bell's Inequality with Cosmic Photons: \\ Closing the Setting-Independence Loophole}

\author{Jason Gallicchio$^{1}$, Andrew S. Friedman$^{2}$, and David I. Kaiser$^{2}$}
\email{Email addresses: gallicchio@uchicago.edu; asf@mit.edu; dikaiser@mit.edu}

\affiliation{$^1$ Kavli Institute for Cosmological Physics, University of Chicago, Chicago, Illinois 60637 USA \\
$^2$ Center for Theoretical Physics and Department of Physics, Massachusetts Institute of Technology, Cambridge, Massachusetts 02139 USA }

\date{\today}

\begin{abstract}
We propose a practical scheme to use photons from causally disconnected cosmic sources to set the detectors in an experimental test of Bell's inequality. In current experiments, with settings determined by quantum random number generators, only a small amount of correlation between detector settings and local hidden variables, established less than a millisecond before each experiment, would suffice to mimic the predictions of quantum mechanics. By setting the detectors using pairs of quasars or patches of the cosmic microwave background, observed violations of Bell's inequality would require any such coordination to have existed for billions of years --- an improvement of 20 orders of magnitude. 
\end{abstract}

\keywords{Bell's Theorem, Setting Independence, Quasars, CMB}
\pacs{03.65.Ud; 42.50.Xa; 98.54.Aj; 98.70.Vc. To Be Published in {\it Phys. Rev. Lett.} 
(2014). }

\maketitle

To date, every published experimental test of Bell's inequality has yielded results compatible with the predictions of quantum mechanics. In light of this robust experimental evidence, Bell's theorem implies that one or more eminently reasonable assumptions about the 
nature of the world must be abandoned or revised \cite{bell64,bell87}. These include locality 
\cite{freedman72,aspect82,weihs98,toner03}, fair sampling of inefficient detectors~\cite{chsh,garg87,eberhard93,giustina13,christensen13}, and detector setting independence (sometimes called freedom-of-choice or free will)~\cite{brans88,conway06,thooft07,scheidl10,hall10,barrett11,hall11,fritz12}. 
While Bell tests are often interpreted as evidence for abandoning the specific assumption of locality to explain the experimental results, relaxing the other assumptions leads to loopholes that could salvage a local realist view where quantum mechanics is incomplete and there are local hidden variables describing its missing degrees of freedom.

Compared to the locality and detector-efficiency loopholes, setting independence has received far less scrutiny, though arguably the standard interpretation of Bell tests is most vulnerable to the setting-independence loophole. Recent calculations have demonstrated that if the setting-independence assumption were false, then rival models could reproduce the predictions of quantum mechanics if the detector settings shared even a small correlation with some local hidden variables \cite{hall10,barrett11,hall11}. 
For example, singlet state correlations in the common two-setting, two-outcome Bell test with entangled photons could be reproduced by a local model that allows as little as 1/22 of a bit of mutual information to be shared between the detectors' polarizer orientations and the hidden variables \cite{hall11}.
This means that a local explanation of observed violations of Bell's inequality could be maintained if only one out of every 22 seemingly ``free choice" binary detector settings were determined by some prior ``conspiracy," established within the shared past light cones of the detectors and the source of entangled particles.  To instead violate Bell's inequality with signaling, a full bit of communication is required~\cite{toner03}.

Performing a loophole-free Bell test and decisively closing the setting-independence loophole remains an important goal not just in the arena of fundamental physics, but in the burgeoning field of quantum information science \cite{scheidl10,giustina13,christensen13}. If hidden variable models of any sort are viable, upcoming quantum encryption schemes could be broken by a sophisticated future eavesdropper that learns to measure the previously ``hidden" variables \cite{barrett11}.

\begin{figure}
\includegraphics[width=0.5\textwidth]{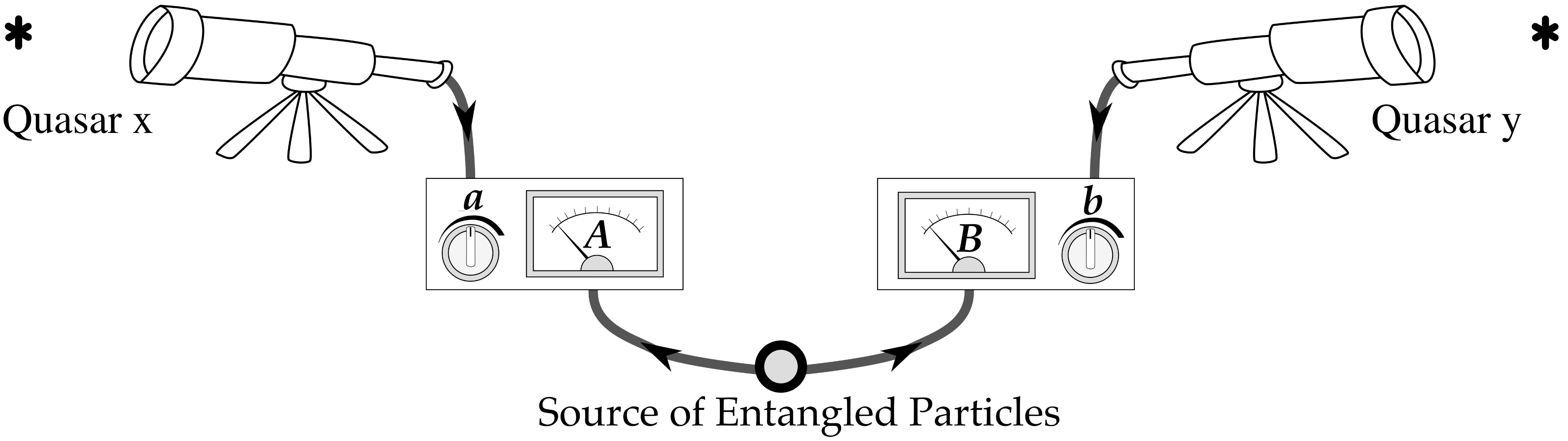}
\vspace{-0.6cm}
\caption{\label{devices}
Schematic of proposed experiment where cosmic sources determine the detector settings in an otherwise standard Bell-type experiment.
\vspace{-0.3cm}
}
\end{figure}

Our proposed ``cosmic Bell" experiment, illustrated in Fig.~\ref{devices}, seeks to narrow this loophole more than any experiment performed to date, using causally disconnected cosmic sources to set the detectors while the entangled pair is in flight. By using cosmic sources that are farther and farther away, we may put quantitative bounds on the distances and timescales over which any such hidden-variable ``conspiracy" must act.  If violations of Bell's inequality are still observed with all other loopholes closed, a local realist explanation would require that the correlations were put in place billions of years ago. Existing state-of-the-art experiments, in contrast, have used quantum random number generators (QRNGs) to set the detectors (e.g. \cite{weihs98,scheidl10}, see \cite{pan12} for  review). The setting-independence loophole in such scenarios requires correlations to have been established merely milliseconds before each detector's measurement, rather than billions of years earlier. Our ``cosmic Bell" experiment would thereby yield an improvement of 20 orders of magnitude.

\begin{figure}  
\includegraphics[width=0.47\textwidth]{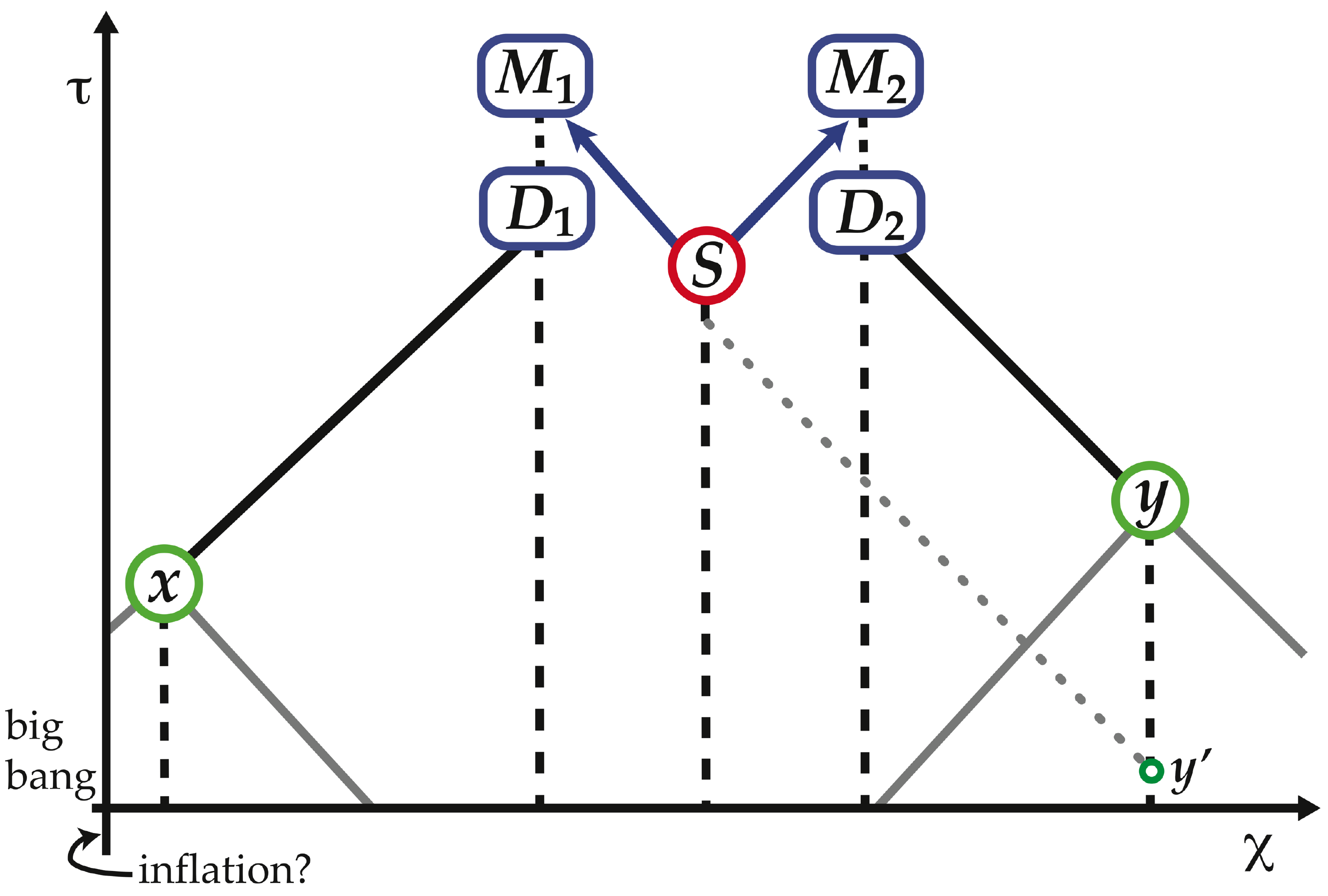}
\vspace{-0.3cm}
\caption{\label{spacetime}
Conformal diagram showing conformal time 
versus comoving distance for the entire history of the visible universe.
In these coordinates, 
null geodesics appear 
as $45^\circ$ diagonals. 
Light from quasar emission events $x$ and $y$ is used to determine the detector settings at events $D_1$ and $D_2$. Meanwhile, spacelike-separated from events $x$, $y$, $D_1$, and $D_2$, the source $S$ emits a pair of entangled particles that are measured at events $M_1$ and $M_2$. 
The quasar emission events can be at different redshifts, provided their past light cones (solid gray lines) share no overlap with each other or with the worldline of the source or detectors since the time of the hot big bang. 
Event $y'$ lies within the past light cones of $y$ and $S$ and can influence both, but not $x$.
\vspace{-0.35cm}
}
\end{figure}

Figure \ref{spacetime} shows a conformal diagram of our setup. If the quasar emission events satisfy the conditions on redshift and angular separation (as viewed from Earth) as detailed in \cite{FKG13}, then their past light cones share no overlap with each other or with the worldline of Earth since the time of the hot big bang, which we take to be the end of post-inflationary reheating, should any period of inflation have occurred in the early universe \cite{guth81,guth05}.
Spacelike separation prevents communication and forces any classical correlations to have been set up in the past light-cone overlap region.

The same basic protocol could be extended to test quantum entanglement in a three-particle GHZ state \cite{GHSZ90,Mermin90b}. A triplet of quasars would be used, each satisfying pairwise constraints on light-cone overlap. Local hidden-variable explanations for GHZ correlations require far greater violation of the setting-independence assumption. In a typical three-particle GHZ test that measures only one of two orthogonal spin (or polarization) bases for each entangled particle, $0.415$ bits rather than $0.046 \simeq  1/22$ bits are required to mimic the quantum expectations~\cite{hall11}.

For either setup, sufficiently distant quasars may be used to push any suspected coordination between detectors to times earlier than the hot big bang.
However, if the source $S$ could somehow tailor its emissions based on partial information about detector settings, causality alone would only require that local hidden variables establish some coordination with events that occurred before the quasars' emission. For example, there could exist an event $y'$ within the past light cones of events $y$ and $S$. If that event determined the properties of the quasar emission, which in turn determined the setting of detector 2, then causality alone would not prevent the source from exploiting information from $y'$ to predict detector 2's setting. Even in such a ``smart source'' scenario, use of distant quasars still pushes $y'$ deep into cosmic history, and would require any such information to be preserved over cosmological distances and times and to be identifiable as pertinent by the source amid all the other data within its past light cone.

{\it Quasars} --- Quasars are the brightest continuous astronomical sources at cosmological distances and have been observed out to high redshifts $z=7.085$ \cite{mortlock11}, farther than the most distant supernova $z=2.357$ \cite{cooke09}.  While some gamma-ray bursts are more distant $z \sim 9.4$ \cite{cucchiara11}, and their optical and IR afterglows can be brighter than comparable redshift quasars \cite{kann10}, such transient sources are difficult to use for our purposes.

\begin{figure} 
\includegraphics[width=0.5\textwidth]{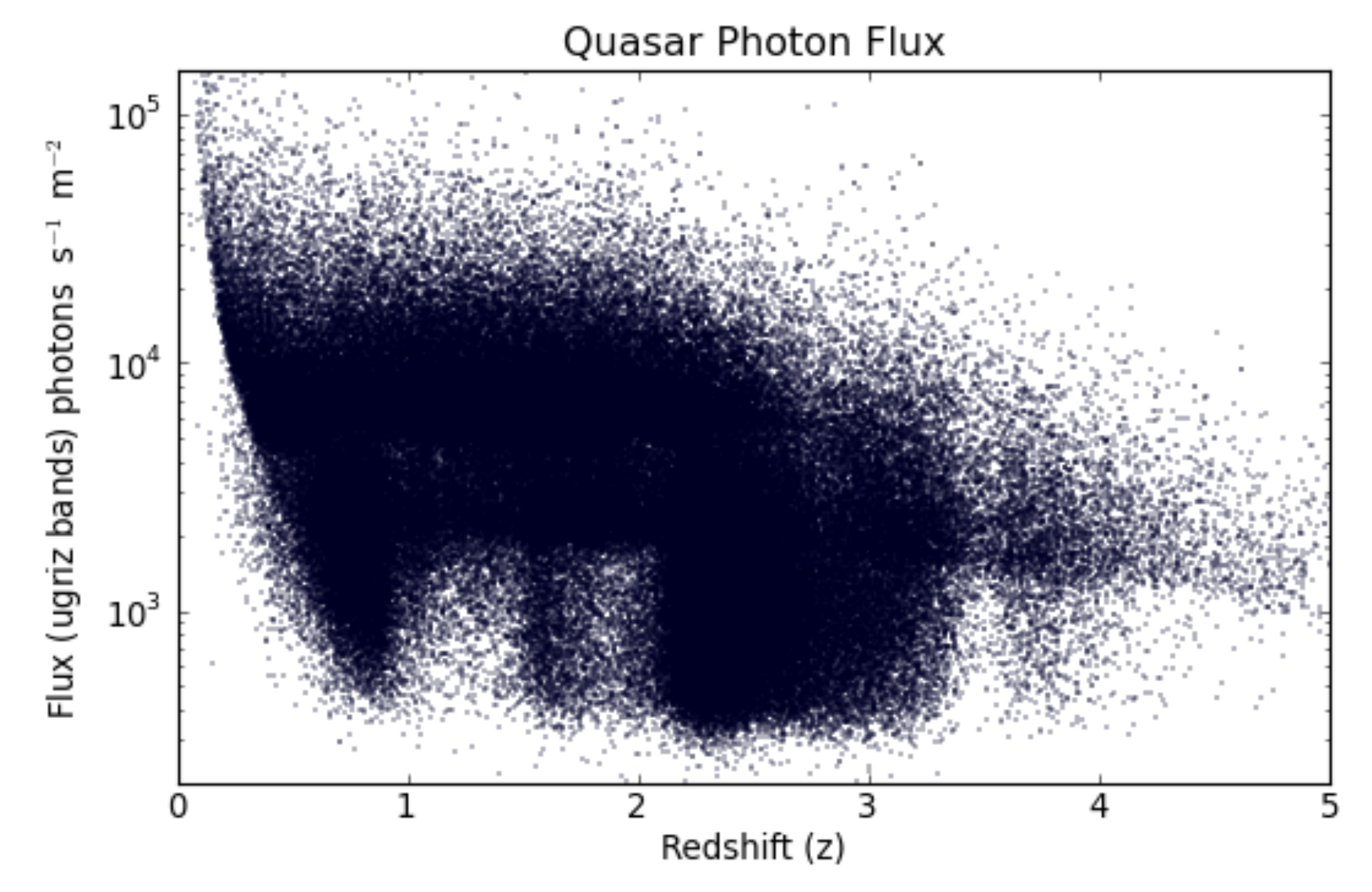}
\vspace{-0.8cm}
\caption{\label{quasarflux}
Optical $ugriz$-band \cite{fukugita96} photon flux from quasars in the Sloan Digital Sky Survey (SDSS) \cite{paris12}. 
Measured photometric brightness in each band was converted to an approximate photon rate and then summed.  Though this is a biased sample, the optical flux of known quasars yields candidate sources and sets the scale for telescope size, distance between the entangled particle source and detectors, and quasar photon coincidence rate.
\vspace{-0.45cm}
}
\end{figure}

Quasars on opposite sides of the sky with redshifts $z > 3.65$ have been causally disconnected from each other and from our worldline since the hot big bang, given best-fit $\Lambda$CDM cosmological parameters from {\it Planck}~\cite{ade13}.
Atmospheric extinction and near-horizon noise require ground-based telescopes to view quasar pairs with separations of less than $180^\circ$. These must be correspondingly farther away to guarantee past causal independence.  Feasible separations and redshifts from \cite{FKG13} are:

{\small 
\begin{center}
\begin{tabular}{l|cc}
 & Angular Separation & \ \ \  Redshift \ \ \  \\
\hline
2-Way Space    &    $180^\circ$     &   $z > 3.65$   \\
2-Way Ground   &    $130^\circ$     &   $z > 4.13$   \\
3-Way Space    &    $120^\circ$     &   $z > 4.37$   \\
3-Way Ground   &    $105^\circ$     &   $z > 4.89$   \\
\end{tabular}
\end{center}
}

Quasars emit most strongly in the rest-frame UV, around the 121.6\,nm Ly-$\alpha$ hydrogen line.  Redshifts of interest move this into the visible and near-IR region. Due to increased IR sky noise, optical photons are most useful for ground-based tests. As shown in Fig.~\ref{quasarflux}, recent surveys like SDSS include substantial optical photon flux from quasars at such redshifts.

To turn quasar light into a bitstream, we can use the arrival time, wavelength, or polarization.  Since accurate timing is already needed to record the entangled particles' arrival, identical time-stamp electronics could be used to record the quasar photons' arrival. Detector settings can be switched based on whether the quasar photon arrived on an even or odd microsecond.
A more sophisticated scheme can get more bits of entropy by whitening the exponential distribution of arrival times~\cite{wahl11,wayne09}.

Using nearby quasars (or stars) would push back causal overlap many orders of magnitude compared to current experiments, but pushing it back to the big bang is also feasible with current technology.
For a given flux $F$ of photons from a quasar, the rate of photons arriving at a telescope is $r = F \pi (d/2)^2$, where $d$ is the telescope diameter. Within a time interval $\Delta t$, and for a detector efficiency $\eta$, the average number of photons detected is $\mu = \eta \, r \Delta t$. Assuming Poisson statistics,
the probability of detecting one or more quasar photons within that period is 
$P=1 - e^{- \eta \, r \Delta t}$.
The probability that both detectors register at least one photon is
\begin{equation}
P_{2} = 
\left[ 1 - e^{-  \eta \, r_1 \Delta t} \right] 
\left[ 1 - e^{-  \eta \, r_2 \Delta t} \right] .
\label{P12}
\end{equation}
If the baselines $L$ between the source of entangled particles and the detectors are sufficiently long, we may ensure that the time required to register the quasar photons (and adjust the detector settings) is shorter than the entangled particles' flight time. For a symmetric arrangement, we therefore take $\Delta t \simeq L / c$. 
For realistic values of 
$d=1$\,m,
$\eta = 0.50$, $L = 50$\,km, and 
$F \sim 2 \times 10^4 \> {\rm s}^{-1} \> {\rm m}^{-2}$ at $z \sim 4.13$ (see Fig.~\ref{quasarflux}), we find
$P_{2} = 0.53$. During about half the experimental runs, both detector settings would be determined by quasar photons. For a ground-based GHZ test, the more distant quasars have about a third the flux. For $L\sim 150$\,km baselines, the triple-coincidence probability is $P_{3} = 0.38$. 
Locality-preserving Bell tests with $L \sim 144$\,km
have already been achieved, as have entangled photon pair production rates of $>10^7$\,Hz  \cite{scheidl10}.
With coincidence rates for both setups of $\sim 10^3$\,Hz,
we could achieve $\sim  10^6$ triggered experimental runs in only 15 minutes. 
Runs in which any or all detectors were not triggered by quasars would serve as useful controls.

The required detector technology also exists.
Superconducting transition edge sensors (TES) have been used to detect entangled photons in Bell tests that close the detector-efficiency loophole~\cite{giustina13,christensen13}. These sensors offer a combination of photon number resolution and detection efficiency larger than $\eta = 97\%$ at 820\,nm~\cite{smith12}, while being virtually free of dark counts~\cite{lita08}.  
Their timing jitter of 78\,ns provides adequate resolution for our long-baseline experiments and could be used for both quasar and entangled-pair detection.
Avalanche photodiodes (APDs) have reduced efficiencies $\eta \sim 50 \%$, but offer much better timing resolution (tens of picoseconds)~\cite{smith12} and have already been used for nanosecond optical astronomy~\cite{howard04}.
Selecting an observing site where the brightest pairs and triplets are well above the horizon for much of the year can maximize the number of experimental runs.
Reducing the telescope area by a factor of 2 would reduce the double-coincidence rate by 4, and the triple-coincidence rate by 8, assuming the quasar signal to noise ratio was still acceptable.  Similarly for decreasing the baseline, given sufficiently fast detector responses.

To rule out local hidden-variable explanations for experimental results, the detector-setting photons must be of genuine cosmic origin. Hence we must also close the ``noise loophole": photons of more local origin from airglow, light pollution, zodaical light, and scattered starlight must be minimized by exposing the detector to a small angular area on the sky. These  backgrounds, along with dark counts from the detector, must be estimated by pointing at a dark patch of sky near the quasar. For a two-particle Bell test, a conservative noise limit is $0.046 \sim 1/22$ of the signal rate~\cite{hall11}. SDSS measures sky glow, and their brightest quasars with $z > 3.65$ only exceed this limit on dark nights, mostly in the $r$ and $i$-bands (623 and 764\,nm). A space-based experiment avoids sky noise from near-IR airglow and could take full advantage of the factor of 2-4 increase in quasar photon flux by including the near-IR $Y\!JHK$ bands.
Noise constraints are an order of magnitude weaker $\sim 0.415$ for a 3-particle GHZ test \cite{hall11} for quasars that are a third as bright.

Photons of cosmic origin should not be altered significantly as they travel through the intergalactic medium, our atmosphere, or the telescopes. All distant photons must at least be affected identically by such media in a way that varies on slow time-scales, like refraction through slowly-varying gas. Away from the plane of the Milky Way, space is indeed transparent. In gamma-ray bursts and supernovae, all the photons arrive at nearly the same time: they are `prompt' or `ballistic,' rather than delayed by some interaction~\cite{cordes98}. More generally, ignoring effects of intervening media is comparable to the assumption made in current Bell experiments, that fiber optics do not significantly alter the properties of entangled photons.


\begin{figure}  
\includegraphics[width=0.5\textwidth]{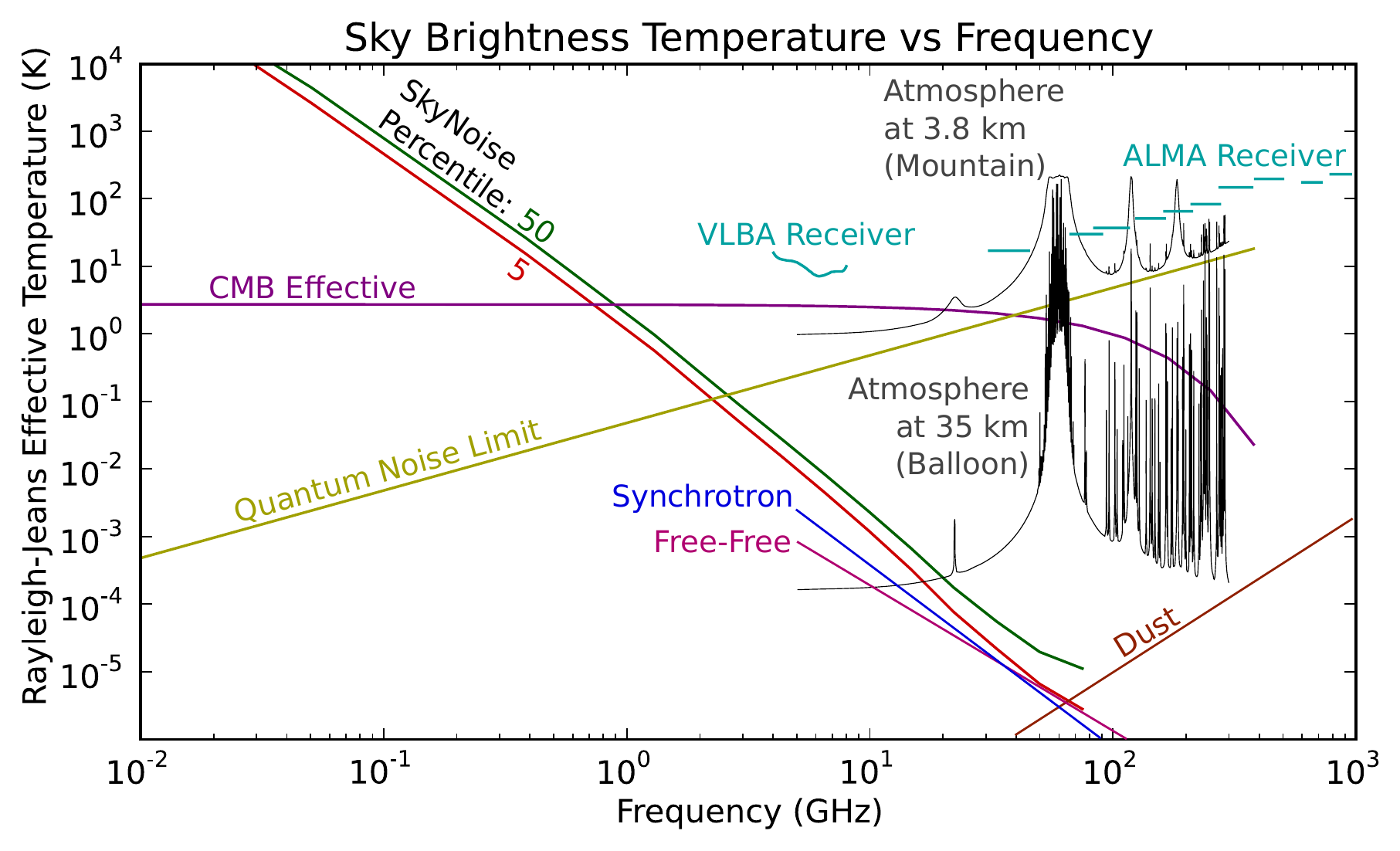}
\vspace{-0.9cm}
\caption{\label{gsm_atmos}
\footnotesize
Effective noise temperature for various sources, including galactic emission from the Global Sky Model \cite{deOliveiraCosta08}, the atmosphere from COFE \cite{leonardi06}, and the noise temperature of state-of-the-art coherent receivers. 
See \cite{hanany13} for a similar figure, along with a photon-noise plot with {\it Planck} detectors. 
\vspace{-0.55cm}
}
\end{figure}

{\it Cosmic Microwave Background} --- The CMB has many appealing features for setting detectors in a causally independent way. CMB patches separated by only $2.3^\circ$ share no causal overlap after the hot big bang \cite{FKG13}, so both receivers can look almost straight up through very little atmosphere. 
There is no need to wait for the brief window during the Earth's rotation when selected quasars are observable through low airmass. 
The bitstream-creating fluctuations would come from the Poisson photon noise of the incoming cosmic radiation. Unfortunately, local noise sources often swamp the instantaneous CMB signal, including the atmosphere, galactic emission, foregrounds, local electromagnetic interference, ground pickup, and detector noise. See Fig.~\ref{gsm_atmos}.

Since the quantum noise limit does not apply, incoherent bolometers have better noise properties than coherent detectors at 80-300\,GHz. Here, galactic emission is low compared to the CMB, but atmospheric emission dominates in any ground-based experiment. Moreover, bolometers operate on thermal timescales of milliseconds, making them problematic to rapidly determine detector settings in a small-baseline experiment. 
Maximizing the instantaneous CMB signal to noise requires atmospheric conditions achievable only from high altitude balloon experiments or satellites positioned hundreds of kilometers from the entangled particle source, each pointing at opposing patches of sky.  The CMB measurements must then be transmitted down to the detectors or the entangled photons must be transmitted up. Ground-to-space entanglement proposals using the International Space Station have already been suggested \cite{scheidl13}. Avoiding light-cone overlap would result in strict latency and positioning requirements.

The detectors' intrinsic (phonon, Johnson, and readout) noise must be reduced below the Poisson photon noise from the CMB.
TES bolometers at 150\,GHz on the EBEX balloon experiment nearly achieve this~\cite{hubmayr09}, as do {\it Planck's} spider web bolometers~\cite{Planck2011HFI}, which are also used on the {\it Archeops} balloon~\cite{Archeops2007}.
Current CMB experiments have no reason further optimize their photon noise limited detectors, and they typically use single-mode optics.  This photon noise is our detector-setting signal, and a multi-mode Winston cone could increase it relative to the intrinsic noise at the cost of wider beams~\cite{hanany13}.  The three-particle GHZ setup is appealing for the CMB because this signal to noise requirement is an order of magnitude less strict than for 2-particle states.  And finding three causally-disconnected spots on the CMB is easy compared to finding three bright quasars that meet the angle and redshift requirements.

{\it Conclusions} --- Until recently, most discussions of Bell tests simply assumed experimenters were able to choose their settings freely. While seemingly quite reasonable, {\it local realism} seemed equally reasonable before Bell's theoretical work \cite{bell64} and the first experiments \cite{freedman72,aspect82}. Recent work \cite{scheidl10,hall10,hall11,barrett11} demonstrates that  the standard interpretation of Bell tests is particularly vulnerable to the setting-independence loophole. Our ``cosmic Bell'' proposal uses the causal structure of space-time to improve the limits on possible correlation between settings and local hidden-variables by 20 orders of magnitude, forcing any ``conspiracy" to have been enacted billions of years ago, rather than milliseconds before a given measurement.

If such an experiment were to be performed, closing all other loopholes, several outcomes are possible. Most likely the Bell inequalities would be violated for every combination of redshifts and angular separations of cosmic sources, regardless of whether the sources' past light cones shared any overlap since the hot big bang. Such results would be in keeping with the predictions of quantum mechanics. In that case, the experiment would have succeeded in closing what is arguably the most crucial outstanding loophole in tests of Bell's inequality. All local hidden-variable theories would be constrained as much as is physically possible in our universe, leaving only super-deterministic cosmic conspiracies, which themselves may not be falsifiable \cite{thooft09}. The usual inferences from Bell tests would then be on as firm a ground as possible. 

An intriguing possibility would be if the degree to which the Bell inequalities were violated showed a statistically significant dependence on the extent to which the past light cones of the cosmic sources overlapped, or how long ago the overlap occurred.
Nearby astronomical sources can probe recent overlap; even by triggering on nearby stars in the galaxy, we could push any conspiracy back 13 orders of magnitude in time, before recorded human history.  Quasars can probe intermediate Hubble-scale overlaps going all the way back to the hot big bang. And the CMB can push this overlap many $e$-foldings back into any inflationary period.  If experimental systematics could not explain such results, and if other experiments confirmed them, perhaps some local hidden-variable theory really were viable and the requisite correlations could be traced to an era of early-universe inflation. In such a scenario, some physical mechanism like inflation would be responsible for establishing the correlations observed in the CMB, as well as correlations between later events like quasar emissions on opposite sides of the observable universe. Such a result would certainly be unexpected, though it would open up the possibility of testing both our most fundamental understanding of non-locality in quantum mechanics, as well as probing various models of the early universe.

\begin{acknowledgments}
{\it Acknowledgements} --- We are grateful to Scott Aaronson, Anthony Aguirre, David Albert, Jacob Barandes, Bruce Bassett, Craig Callender, Shelly Goldstein, Alan Guth, Ned Hall, Brian Keating, Robert Kirshner, Kamson Lai, Seth Lloyd, Felipe Pedreros, Matthew Pusey, Katelin Schutz, Robert Simcoe, Chris Smeenk, Jack Steiner, Nicholas Stone, Leonard Susskind, Max Tegmark, David Wallace, and Anton Zeilinger for stimulating discussions. 
This work was supported in part by the U.S. Department of Energy (DOE) under Contract DE-FG02-05ER41360. ASF was also supported by the U.S. National Science Foundation (NSF) under Grant SES-1056580. JRG is supported by NSF grant ANT-0638937.
\end{acknowledgments}

\vspace{-0.15cm}

{\it Note added.} --- Recently, we discovered that others have briefly mentioned the basic premise of using cosmic sources to determine detector settings \cite{maudlin94,scheidl10,zeilinger2010dance}. However, we believe we are the first to develop a realistic protocol for such an experiment, calculating appropriate causal conditions \cite{FKG13} and quantifying basic detector requirements for real candidate sources in our universe.

\bibliography{./selection,./sn,./sngroup,./jasonDK4}

\end{document}